# SDN Intrusion Detection Using Machine Learning Method

Muhammad Zawad Mahmud[1], Shahran Rahman Alve[1], Samiha Islam[1] and Mohammad Monirujjaman Khan[1]

[1] Department of Electrical and Computer Engineering, North South University, Dhaka-1229, Bangladesh
Email: {zawad.mahmud1, shahran.alve, samiha.islam2, monirujjaman.khan}@northsouth.edu

## ABSTRACT

*Software-defined network (SDN) is a new approach that allows network control to become directly programmable, and the underlying infrastructure can be abstracted from applications and network services. Control plane). When it comes to security, the centralization that this demands is ripe for a variety of cyber threats that are not typically seen in other network architectures. The authors in this research developed a novel machine-learning method to capture infections in networks. We applied the classifier to the UNSW-NB 15 intrusion detection benchmark and trained a model with this data. Random Forest and Decision Tree are classifiers used to assess with Gradient Boosting and AdaBoost. Out of these best-performing models was Gradient Boosting with an accuracy, recall, and F1 score of 99.87%,100%, and 99.85%, respectively, which makes it reliable in the detection of intrusions for SDN networks. The second best-performing classifier was also a Random Forest with 99.38% of accuracy, followed by Ada Boost and Decision Tree. The research shows that the reason that Gradient Boosting is so effective in this task is that it combines weak learners and creates a strong ensemble model that can predict if traffic belongs to a normal or malicious one with high accuracy. This paper indicates that the GBDT-IDS model is able to improve network security significantly and has better features in terms of both real-time detection accuracy and low false positive rates. In future work, we will integrate this model into live SDN space to observe its application and scalability. This research serves as an initial base on which one can make further strides forward to enhance security in SDN using ML techniques and have more secure, resilient networks.*

## KEYWORDS

*SDN, machine learning; classifier, accuracy;*

## 1. INTRODUCTION

Due to the tagging part, which demands a human output to decide whether an end-product is anomalous or good, detecting anomalies comes across as difficult even with labeled datasets. In order to keep up with the activity of these massive production systems, they must be monitored every ten milliseconds – this results in a flood of data and what we call the Industrial Internet of Things (IIOT). This gets worse, and no manufacturer wants an outlier product. Therefore, the abnormalities are completely unproportioned — which h makes it a needle in a haystack scenario. The rapid evolution in technology, e.g., the Internet and the IoT (Internet of Things) devices, as well as communication systems, imply that hackers are gaining skills at an even faster rate. These kinds of criminals are permanently interested in bank protection, and they have reactive heads to conquer any site that is under demanded against enemy viruses. Consequently, network security is closely related to intrusion detection systems (IDSs).

One of the best solutions in the intrusion detection research community during recent years is machine learning (ML)-based invasion detectors. ML-It enables computers to learn from and improve based on experience. In other words, so long as the computer-based application uses a machine learning model, explicit engineering (programming) isn't really required. It is intelligent — it can learn by itself [1]. Machine learning as a whole is split into two fields of thought: Supervised machine learning and unsupervised machine learning. Learning Here, we have trained models on labeled data, which is shown via supervised machine learning [2]. Unsupervised Machine Learning: Unsupervised machine-learning models are trained on unstructured text data [3].

The objective of this study is to explore supervised machine learning techniques, particularly for binary and multiclass classification. The classification operation happens when a supervised ML model is introduced with a value to predict as discrete [4]. For the latter, sample sizes are large or very large with high-dimensional feature space used to train models. This complexity results in relatively slow training and testing of supervised models. Therefore, feature engineering approaches are indispensable for the dimension reduction of features before training-test stages. This paper contributes by using the supervised machine learning algorithms to make IDS, Random Forest, Decision Tree, Gradient Boost, and Ada Boost, which were experienced in doing so.

## 2. RELATED WORKS

Software Defined Networking (SDN) has brought a significant change in network management with unprecedented flexibility and control. Still, it also provides security challenges that must be dealt with integrally at the SDN Control. In this paper, the authors investigate the potential advances in related attack detection methods using advanced machine-learning techniques to improve Intrusion Detection Systems (IDS) that help improve SDN protection against rather soon sophisticated attacks. SDNs that are evolving have greatly helped enhance and make cloud computing technologies; therefore, they are more scalable, cost-effective, and manageable. This evolution, in turn, creates more security holes, which is why OT networks are arguably less secure than traditional [5]. A different way to avoid these vulnerabilities is by utilizing machine learning representations enclosed in SDN infrastructure, which will be equipped for capturing such gaps from malicious traffic and have been considered with fantastic curiosity. In their research, the authors presented a large benchmark analysis with the NSL-KDD dataset, which is an improved version of KDDCUP'99; however, we have some limitations. They identify these as core difficulties with classical machine learning methodologies, and a systematic investigation such as theirs can provide not only regional benchmarks to compare against but, more importantly, lay the foundation for further research in building alternative intrusion detection frameworks by analyzing data non-linearity characteristics and how well different ML approaches work under those conditions.

Recently, the flexible nature of SDNs has opened them up for numerous attack types, which IRS still comprises with no successful bid to mitigate; hence, lots of attention in research circles has been garnered towards the integration between these approaches and methods namely machine learning. In this paper, an attempt is made to address these vulnerabilities; they came up with a new IDS model that incorporates some best practices like a Hybrid Feature Selection Algorithm (HFS) mingles Correlation-Based Feature Selection (CFS), and Random Forest Recursive Feature Elimination with the use of Light Gradient Boosting Machine (Light GBM in detecting, classifying threats. Their research, illustrated on the NSL-KDD dataset, reveals improved detection metrics, including accuracy, precision-recall, and F-measure compared to existing methods. It also points out the positive aspects of combined machine learning techniques for improving SDN security.

Khanal et al. [6] suggested that due to the many complexities involved in managing traditional distributed network systems, SDNs proposed this separation. Networking would be more convenient if the data/planes were to take place separately from QoS planes. Still, SDNs have specific weaknesses, mainly in their control plane, which, if hacked, will cause the entire network to collapse. The only way to address these security issues is by using an IDS. In this context, their research was developed with the aim of filling a gap in IDS solutions for SDN environments by creating a NITSDN dataset. This work underlines the importance of this dataset by producing a Decision Tree classifier accuracy that achieves up to 99.48%, promising new conceivable ways machine learning can potentially enhance SDN security.

One of the most prevalent forms of network virtualization is SDN, which has revolutionized management while introducing new threat vectors, such as Distributed Denial-of-Service attacks meant to degrade network-level services. Gupta and Grover [7] have conducted a critical comparison of different machine learning algorithms to be used for DDoS detection in SDN. A separate extensive review identifies how deep convolutional neural networks outperform previous models for detecting such threats, achieving an accuracy rate greater than 99.45%.

The emerging trends in network security bring out the importance of effective Intrusion Detection Systems to protect computer networks and information systems. The research by Hassan et al. [8] also brings the incorporation of Convolutional Neural Networks (CNNs) in implementing IDS with a special tune for SDN. This work utilizes a unique dataset, In SDN, which is about to create attack-specific scenarios in SDNs.

Arthi et al. [9] emphasized an innovative SDN-IoT-based intrusion detection system for healthcare using deep-learning and support vector machines to improve the quality of care by enhancing operation safety. This work emphasizes the importance of secure and reliable IoT-based healthcare services at times like a pandemic such as COVID-19. The authors evaluate their framework's effectiveness in detecting and mitigating network intrusions over a Mininet-based emulation with high-performance metrics such as F1 score, precision, and recall. This work adds to the emerging area of securing SDN-enabled IoT networks, especially in healthcare, where cyberattacks could entail life-threatening outcomes.

Isa and Mhamdi [10], in their paper titled "Native SDN Intrusion Detection using Machine Learning," discuss the challenge of securing SDN environments by introducing a native autoencoder-based approach to intrusion detection named Random Forest IDS. The system is built to have a minimal latency and low detection rate while still achieving a high accuracy of mean of 90.19 % using autoencoder only; if combined with a random forest classifier, it scored an accuracy up to 98.4 %. The paper proves that the proposed solution balances performance vs. security and also has the lowest impact on overall Field Programmable Gate Array (FPGA) based SDN controller throughput and latency.

In SDN, the recovery speed from a network flow incursion represents critical elements to maintain functions and security within networks. For this challenge, Hammad et al. [11] presented the Machine Learning Network Intrusion Recovery (MLBNIR), which complements the recovery of intrusions in SDN by selecting a backup path strategically according to traffic patterns. Trained on a specialized SDN dataset, their system can cut recovery time from an intrusion by as much as 90% while saving up to 57% of network bandwidth over existing methods. Their research results show that the MLBNIR approach can reduce detection time and maximize bandwidth, combining low recovery latency with high efficiency, representing a promising solution for intrusion response.

Ahmad et al. [12] focus on the applicability of Machine Learning solutions to address these security problems within SDN environments. Their research work considers a number of ML algorithms, such as SVM, NB, and Decision trees, along with Logistic Regression under a practical environment where the SDN controller is targeted by DDoS attacks. These results showed that the SVM performs better than other algorithms with 97.5% accuracy and 97% precision, but all four other algorithms show more accuracy than Logistic Regression >96%. These results highlight the capability of ML-enabled approaches to further secure future SDN deployments.

Chaganti et al. [13] proposed a novel deep learning-based intrusion detection paradigm utilizing a Long Short-Term Memory (LSTM) framework specifically designed for SDN-enabled IoT networks. The results obtained from their work showcased the capability of the LSTM model to classify and detect various assaults that took place in a network with an accuracy of 0.971, which surpasses other deep learning classifiers, i.e., DNN as well as CNN, even its traditional machine learning counterpart like SVM, was unable to outperform it.

Hadem et al. [14] implemented an SDN-based Intrusion Detection System, which employs Support Vector Machines in combination with Selective Logging for IP traceback. Their method reaches up to 95.98% detection on the complete NSL-KDD data set and 87.74% over some essential features used by other authors, showing strong performance while remaining efficient in terms of computation usage.

Shaji et al. [15] emphasized a smart IDS called SD-IIDS, which utilizes ensemble ML models to detect DDoS attacks in the realm of software-defined networking (SDN). The authors have devised two ensemble models, namely SVC-RF and RF-LR. The Support Vector Classifier bagged with a Random Forest (SVC) classifier and the Random Forest classifier are then ensembled together, and the Random Forest is combined with Logistic Regression; these hybrid classifiers perform very well for low-level network flow categorization. The binary RF-LR model gained an accuracy rate of 99.79% with a reduced False Alarm Rate (FAR), and the multi-class RF-LR netted a top-1 test set prediction at 99.54%, besting all other models!

SDN technologies brought a new era of network architecture as well as increased the level of security mechanism that is exposed to cyber-attacks. In order to get over these gaps in SDN, Hadi and Mohammed [16] presented a deep learning-based Network Intrusion Detection System, abbreviated NIDS-DL, that is implemented for the first time this way. In addition, they used various deep learning classifiers (CNN, DNNs /RNN/ LSTMs and GRUs) with the same NIDS framework that offer a more enhanced feature selection technique from NSL-KDD dataset of which apparently some tremendous outcomes have been illustrated in terms of perfect accuracy as well precision followed by better F1-score where CNN tops even all others. (add accuracy).

The research paper by Houda et al. [17], entitled "A Novel Machine Learning Framework for Advanced Attack Detection using SDN", deals with some vital challenges in this field. The authors have presented a novel multi-module machine learning framework based on unsupervised learning for quickly and efficiently detecting network security threats. They have showed that their approach drastically improves accuracy and detection rates with lower computational costs compared to state-of-the-art methods applied to the UNSW NB15 dataset of network security. (add best model and accuracy)

In this investigation, we utilized five machine learning algorithms, and all of them were accurate to better than 99%. With the Gradient Boosting classifier, we achieved the highest accuracy of 99.87%. As a consequence, we were able to reduce the accuracy gap in our study. The use of these four models has the benefit of allowing for comparative examination. These comparisons help us figure out which model has the highest degree of accuracy. Any company may use our technology to identify intrusions by simply implementing it. It will save a lot of personal information, privacy, and money for any firm. The following is the rest of the article: Section 3 describes the experimental approach and procedures, whereas Section 4 analyzes the results, and Section 5 discusses the conclusion.

## 3. METHODOLOGY

Some of the information in this part is very specific. It includes a lot of information about the dataset, the proposed system, and how the study was done.

### 3.1 Proposed System

The block diagram of the suggested system is depicted in Figure 1.

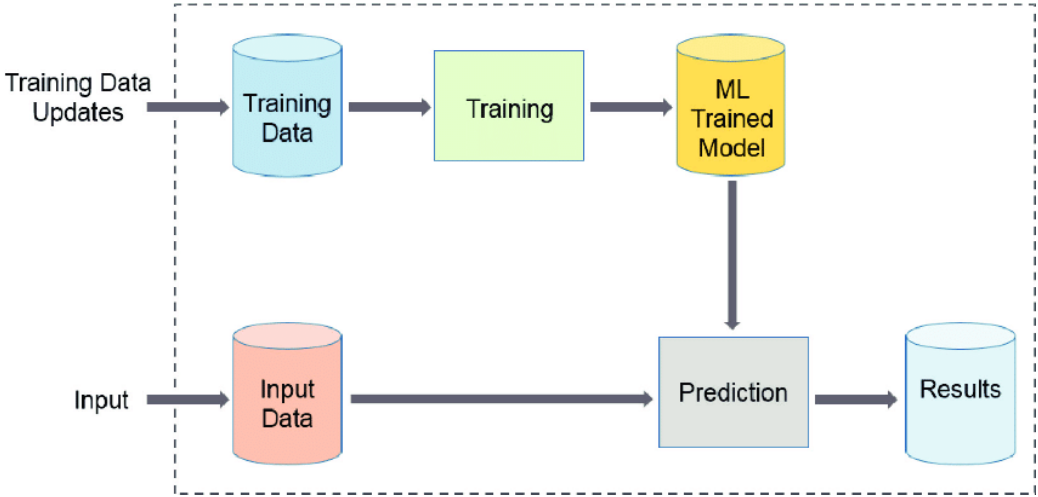

Figure 1: Block Diagram of the Proposed System

### 3.2 Dataset

The experimental approaches are based on the SDN Anomalies Detection dataset [18]. There are a total of 1763 count columns in the cleaned dataset, with the final set restraining itself to a leaner version residing at 1558 features. Class (0 or 1)—whether the product belongs to a bad class or not. The total number of anomalies and normal data is shown in Figure 2. 0 represents normal and 1 data points, respectively.

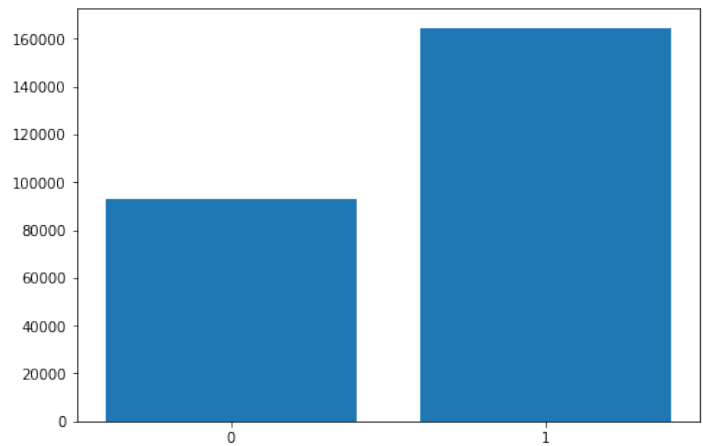

Figure 2: Total number of Anomalies and Normal Data

### 3.3 Data Preprocessing

Raw data is converted into a suitable format (raw input) so that it can be easily fed into the model (or pipeline). It is the first and most crucial step you take while building a machine-learning model. Some common problems in real-world data are noise, missing values, and null values, as well as issues such as the format of the velocity not being suitable for ML models. Data preparation procedures, which involve cleaning and organizing our data to feed it into the model, will, in some cases, help achieve better accuracy levels and speed up on some occasions while building machine learning models. In the event that the dataset has missing data, it can seriously wreck our machine-learning model. Therefore, it is essential to handle the missing values in the dataset and complete those null values. Check null and missing values in the dataset.

### 3.4 Proposed Algorithms

Anomaly detection systems for networks are at the core of network security. Aside from detecting abnormalities, network anomaly detection systems continually monitor and assess the events inside a network. Using a publicly accessible dataset, the researchers tested four machine-learning algorithms for identifying abnormalities. The following are the details:

- Random Forest
- Decision Tree
- Gradient Boosting
- Ada Boost

#### 3.4.1 Random Forest

The Random Forest Classifier was taken as the first algorithm. The Random Forest (RF) has many decision trees, and a single tree area is called a base-classifier in RF language, which is trained on bootstrapped training sample data. All of these three results are then voted to be utilized as estimations. The last prediction is made by a majority vote among the votes of all RF classifiers. Figure 3 shows the RF classifier's block diagram.

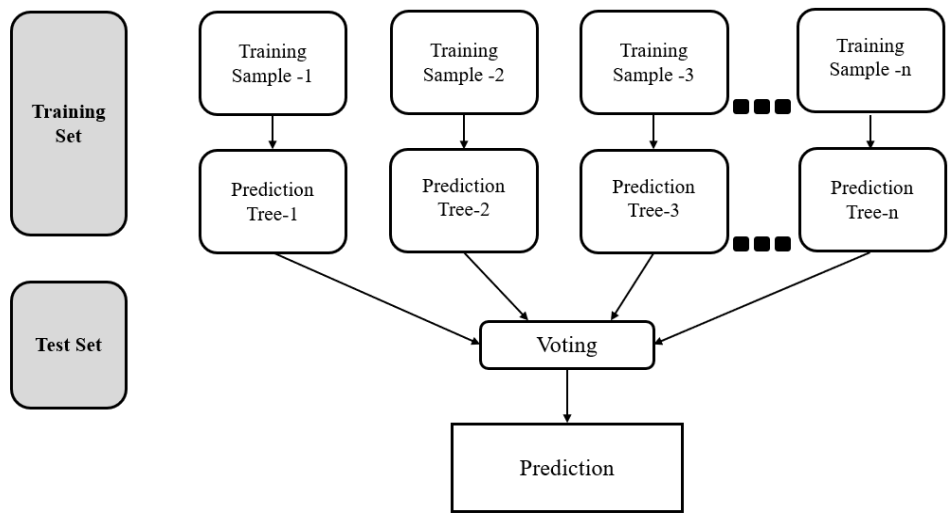

Figure 3: Diagram of Random Forest Classifier

It gives good results because RF gets trained much faster than other algorithms. In practice, the accuracy of SVM is not much influenced by using huge datasets (as it provides a lot of accuracy with big data). In a limited data time series model, Precision has the least impact. Equation 1 shows the basic classification formula of the random forest algorithm.

$$Gini = 1 - \sum_{i=1}^{C} (p_i)^2$$

Equation 1: Classification Formula for Random Forest

### 3.4.2 Decision Tree

Decision tree methods are used in machine learning to solve classification and problems. The root node creates two types of nodes: a.) internal node and b.) leaf node. In branches are the internal nodes, which are also called decision-makers, and at last, output leaves since no more branching is possible. Therefore, it changes the structure of the tree. The core architecture of the decision tree classifier is shown in Figure 4.

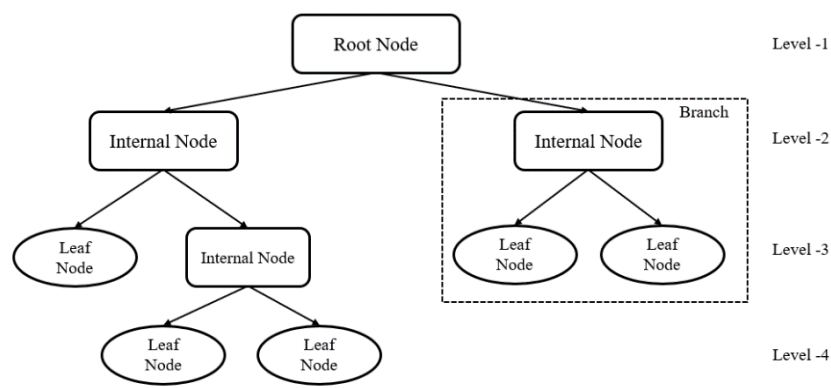

Figure 4: Diagram of Decision Tree Classifier

Most of us have data cleansing forgotten when it comes to DT. (Because what is that bonus ) Well, this one is sort of obvious because it's in tree form. Explaining the DT is easy because it explores how a person goes step by in real-world decision-making. Similarly, the DT classification formula is given in equation 2.

$$IG(D_p, f) = I(D_p) - \frac{N_{left}}{N} I(D_{left}) - \frac{N_{right}}{N} I(D_{right})$$

Equation 2: DT classification formula

### 3.4.3 Gradient Boosting

One of the devolved machine learning methods is called the gradient boosting approach [19]. GBA has been one of the most successful in various challenges. A way for the predictors to tweak their performance based on trial and error. Nevertheless, the amazing feature of gradient boosting is that instead of fitting a classifier to minimize our loss on each step, it fits another model into regression generated from the previous predictor [20]. The schematic design of the GB algorithm is illustrated in Figure 5. Preprocessing of data is not a prerequisite in GBM. That is why it is capable of dealing with missing data. This approach is very versatile and may be used to enhance a variety of different loss functions.

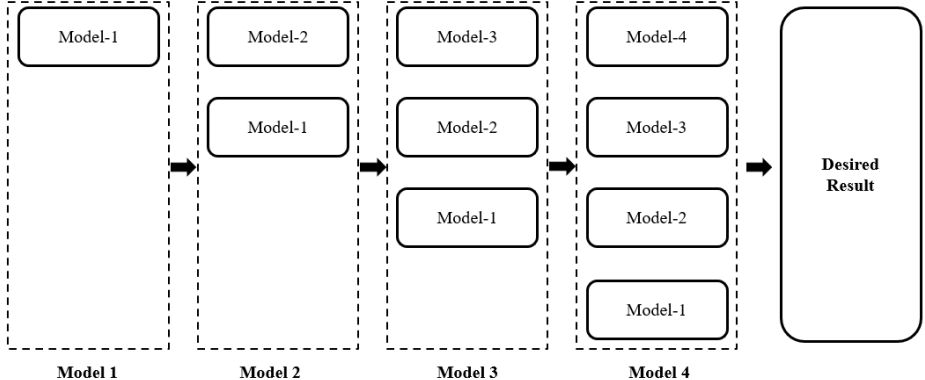

Figure 5: Diagram of Gradient Boosting Classifier

### 3.4.4 Ada Boost

Ada Boost algorithm [21] is a boosting strategy employed in machine learning as an ensemble method. In this process, the weights are reassigned to each instance, where a higher weight (β) is assigned to misclassifying instances [22]. All subsequent learners, except the very first one, are built on top of the previous ones. In other words, weak learners turn into strong ones. Figure 6 shows the conceptual image of how the algorithm constructs an initial model and finds faults in this first created model. Then, the miscategorized record is used as input for the next models. This process goes on until the condition specified is satisfied. It appears that three models are built by providing the errors from the previous model, which is used to boost work.

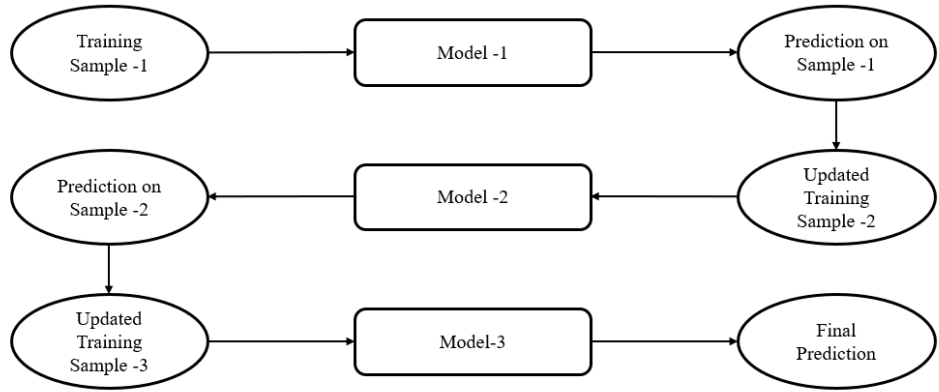

Figure 6: Diagram of Ada Boost Classifier

### 3.5 Evaluation Matrix

The evaluation matrix is a metric that reflects how well ML and DL algorithms execute in terms of the confusion matrix. The confusion matrix will be used to assess the whole list of models. The confusion matrix shows how often our models produce correct and incorrect estimates. This can be seen in Figure 7, where false positives and negatives will go to badly protected values while actual positives and negatives are accurately predicted. Accuracy, Precision-Recall Tradeoff, and Accuracy-recall tradeoff were thus examined to determine the performance of the algorithm aggregating all estimated values in the matrix.

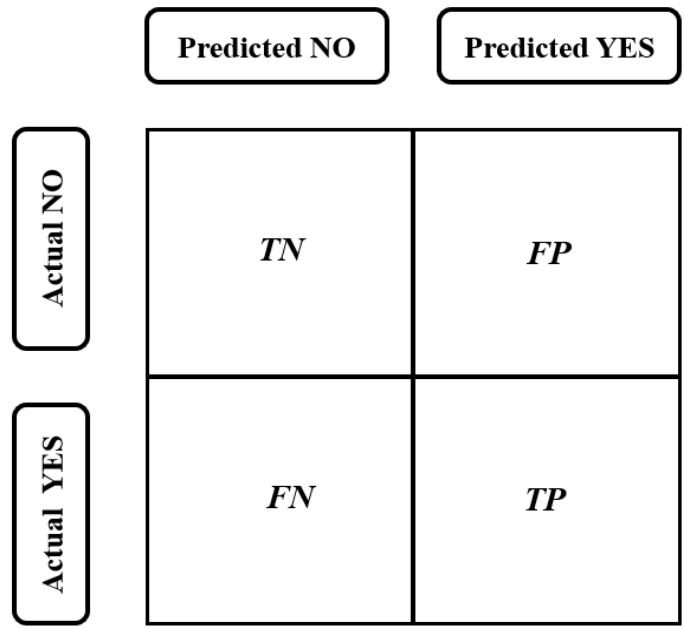

Figure 7: Block Diagram of Evaluation Matrix

## 4. RESULT ANALYSIS

### 4.1 Visualization of Feature Selection

The feature selection technique is shown in Figure 8. Feature selection aids in comprehending how characteristics interact and are connected to one another. It also shows the positive and negative correlations among various features. During the training and testing phase, we only took the features that were positively correlated. All negatively correlated features were dropped out during the preprocessing phase.

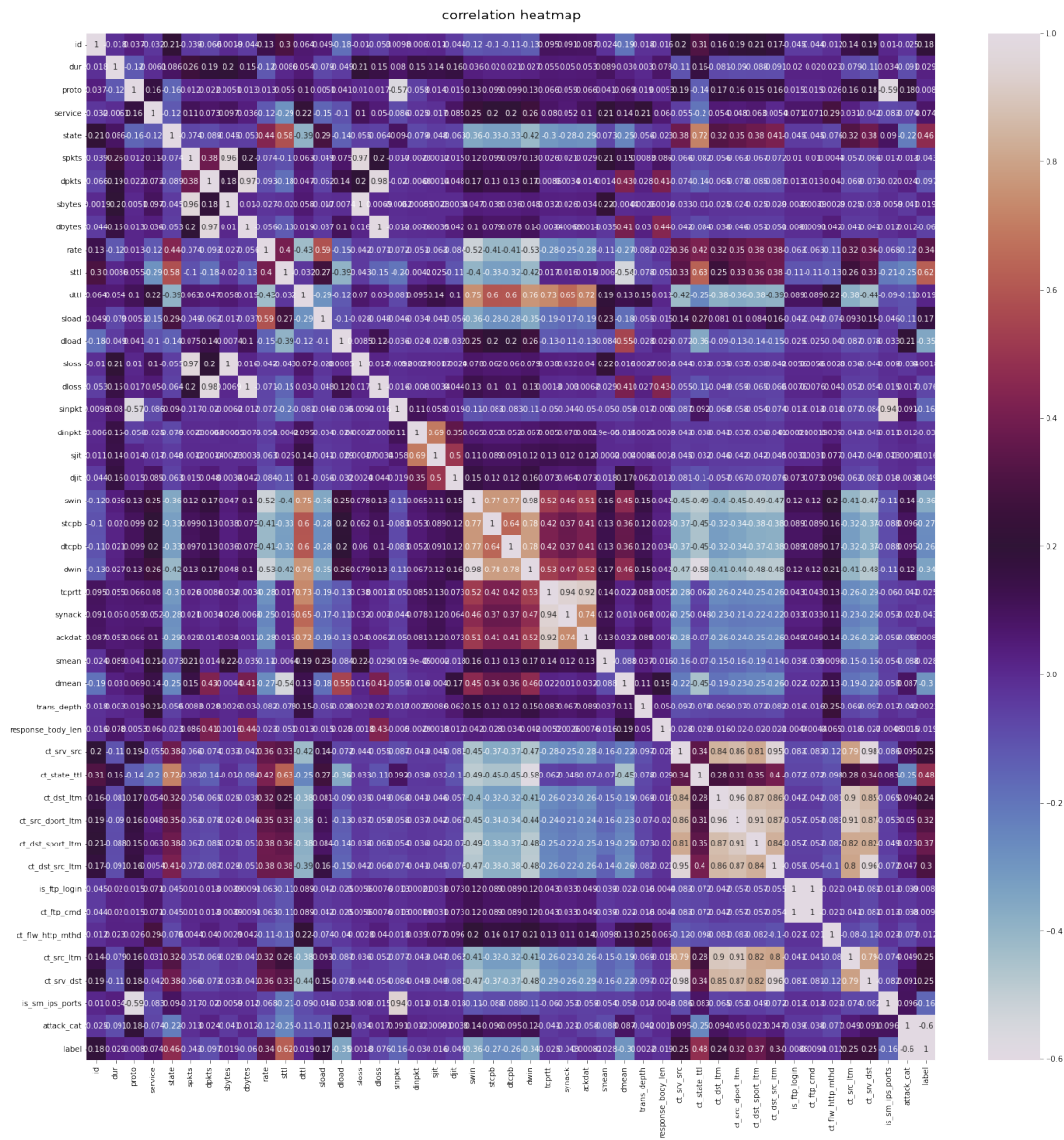

Figure 8: Heat map of Features correlation.

## 4.2 Confusion Matrix and ROC

In this section, the best two models and the worst model's confusion matrix and ROC curve are displayed at first. Later, the applied models of this study are compared in terms of precision, recall, f1 score, and accuracy. Finally, it was compared with the existing ones available for this study.

### 4.2.1 Gradient Boosting

In Figure 9, the confusion matrix of Gradient Boosting is shown. A confusion matrix helps to make comparisons among the values that are achieved through model training. Various values such as true positive, true negative, false positive, and false negative will be used to determine the result and performance. In this confusion matrix, there are 60202 correct positive responses, 62737 correct negative responses, 0 incorrect negative responses, and 187 incorrect positive responses.

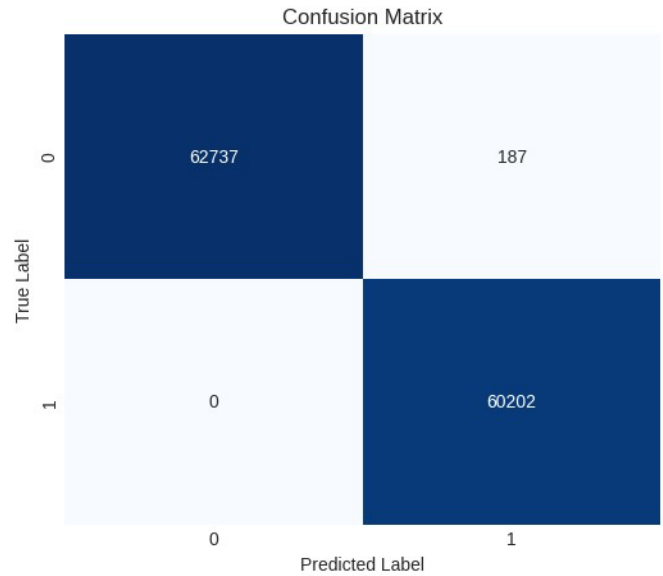

Figure 9: Confusion matrix of the Gradient Boosting classifier

The ROC curve for the Gradient Boosting classifier is shown in Figure 10. The ROC curve of the Gradient Boosting classifier also shows a fantastic model performance with an AUC of 1.00. The sharp rise of the curve to the up-left side clearly indicates that both classes (positive and negative) are almost perfectly separated, so we can get a very high accuracy here with rare cases of false positives or false negatives.

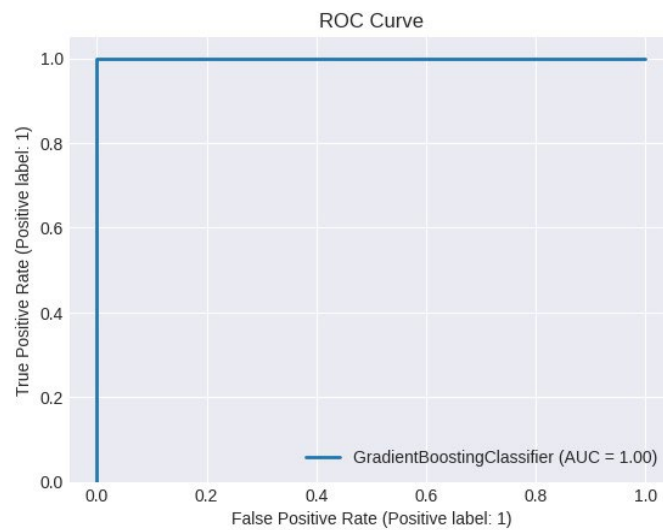

Figure 10: ROC curve of the Gradient Boosting classifier

### 4.2.2 Random Forest

Figure 11 shows the confusion matrix of the random forest algorithm. The prediction made by the random forest model is seen in Figure 11. The confusion matrix depicts the projected result and the model's computed performance. There are 59270 correct positive responses, 63094 correct negative responses, 608 incorrect negative responses, and 154 incorrect positive responses.

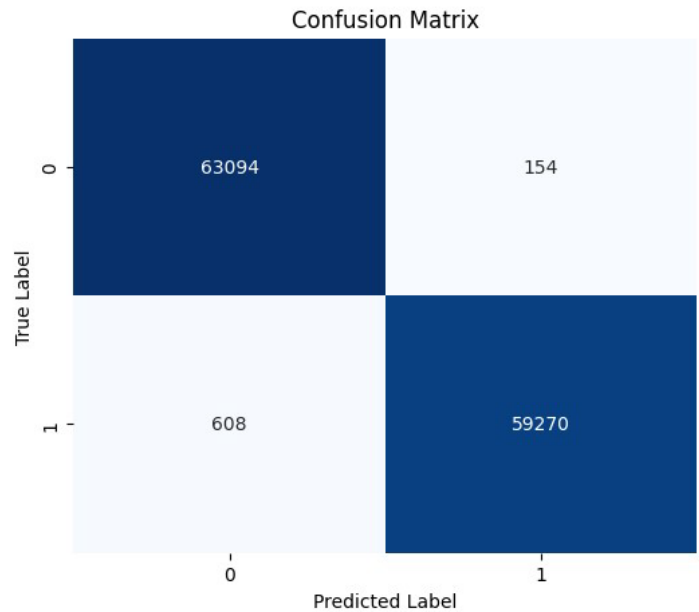

Figure 11: Confusion matrix of the Random Forest classifier

The ROC curve of the Random Forest classifier appears in Figure 12. The ROC curve illustrates that the Random Forest classifier generalizes almost perfectly, having an Area Under the Curve (AUC) of 1.00. The curve is quite close to the top-left corner, showing that our model has a good capability of discriminating between positive and negative classes without any kinds of false positives or negatives.

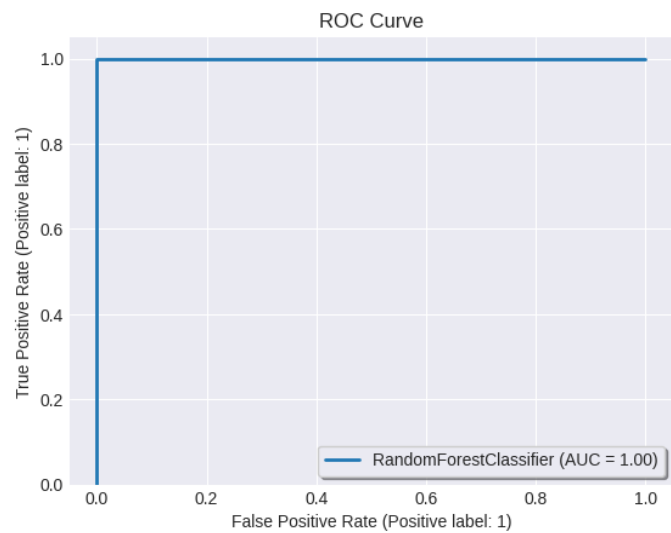

Figure 12: ROC curve of the random forest classifier

### 4.2.3 Decision Tree

The prediction made by the decision tree model is seen in Figure 13. The confusion matrix depicts the projected result and the model's computed performance. There were 122201 correct predictions and 925 incorrect ones.

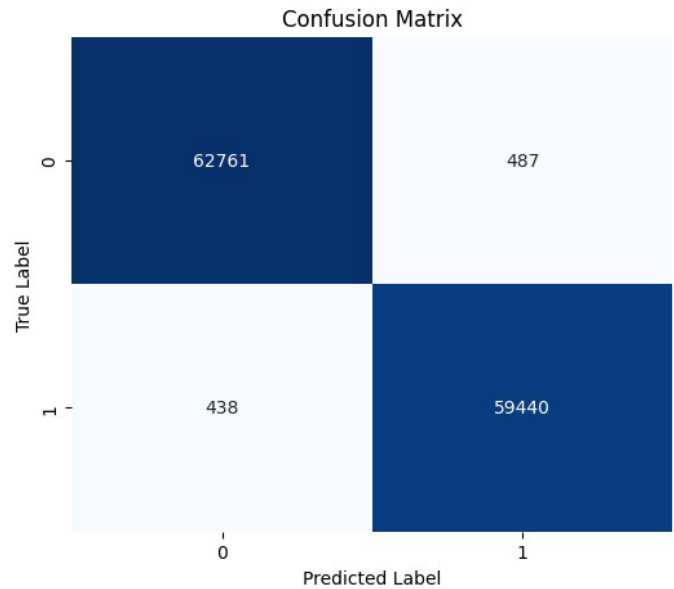

Figure 13: Confusion matrix of the Decision Tree classifier

Figure 14 represents the ROC curve of the Decision Tree classifier. The ROC curve to the Decision Tree classifier shows a performance of 0.99 on AUC, which means that it performed very well as a model. The curve is close to the top-left corner, meaning that it classifies with an optimal balance between sensibility and specificity (i.e., few false positives or negatives).

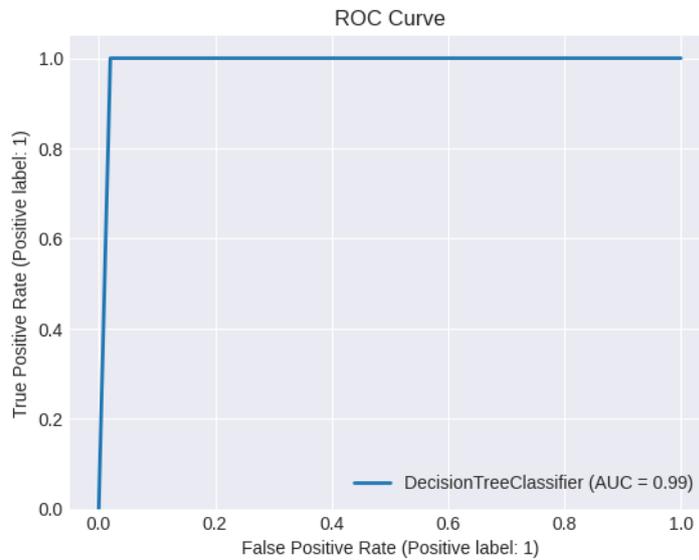

Figure 14: ROC curve of the decision tree classifier

### 4.3 Model Evaluation

Table 1 displays the performance metrics of all the models applied in this study. Among these, the Random Forest achieved the most precision, 0.9974. However, Gradient Boosting was able to outplay all in terms of recall, f1 score and accuracy. Gradient Boosting achieved a remarkable 100% recall. 99.85% f1 score and 99.87% accuracy.

Table 1. Performance Metrics Comparison

| Algorithm Applied | Precision | Recall | F1 score | Accuracy (%) |
|---|---|---|---|---|
| Gradient Boosting | **0.9969** | **1.0000** | **0.9985** | **99.87** |
| Random Forest | 0.9974 | 0.9899 | 0.9936 | 99.38 |
| Ada Boost | 0.9934 | 0.9932 | 0.9933 | 99.34 |
| Decision Tree | 0.9919 | 0.9927 | 0.9923 | 99.24 |

### 4.4 Model Comparison

The comparison between our model and other models that have been studied is shown in Table 3. The gradient boosting clearly possesses the highest accuracy rate of any model in this instance.

Table 3. Comparison of performance results

| Study | Best Model | Accuracy (%) | F1 Score (%) |
|---|---|---|---|
| This paper | **Gradient Boosting** | **99.87** | **99.85** |
| [15] | Random Forest+Logistic Regression | 99.79 | 99.78 |
| [7] | CNN | 99.45 | 99.56 |
| [11] | Random Forest | 98.40 | 98.45 |

## 5. CONCLUSIONS

This paper investigated the performance of different machine learning classifiers for intrusion detection in Software-Defined Networking (SDN) over the UNSW-NB15 dataset. Out of all four evaluated models (Gradient Boosting, Random Forest, Decision Tree, and Ada Boost), Gradient Boosting performed the best by massive margins in all metrics with an upgraded accuracy rate from previous existing models to 99.87 %, recall 100% and f1 score 99.85%. The model's better performance indicates its ability to differentiate normal traffic from malicious flow effectively, which benefits the intrusion detection solution in SDN. The Random Forest model was reliable as well, with 99.38% accuracy. The results demonstrate that ensemble methods, especially Gradient Boosting, can effectively provide a good IDS model for SDN. This paper suggests that the application of new machine learning models can significantly increase network security and, in turn, reduce vulnerabilities in real-time threat detection. In continuation, future research must concentrate on integrating this model in real SDN environments and verifying end-to-end scalability and deployment practicality to ensure more secure network infrastructures.

## ACKNOWLEDGEMENTS

The authors would like to acknowledge North South University for allowing them to use computational resources for this study.